\documentclass[%
 reprint,
 amsmath,amssymb,
 aps,
 longbibliography,
]{revtex4-1}
\usepackage{graphicx}
\usepackage{textcomp}
\usepackage[Symbol]{upgreek}
\usepackage{amsmath}
\usepackage{amssymb}
\usepackage{mathrsfs}
\usepackage{hyperref}
\usepackage{graphicx}
\usepackage{dcolumn}
\usepackage{bm}

\usepackage{xcolor}
\newcommand{\YS}[1]{\textcolor{black}{ #1}}

\begin{document}
	
\preprint{APS/123-QED}
	
    \title{\YS{Divergence-degenerated spatial multiplexing towards ultrahigh capacity, low bit-error-rate optical communications}}
	
\author{Zhensong Wan$^{1,2}$, Yijie Shen$^{3}$, Zhaoyang Wang$^{1,2}$, Zijian Shi$^{1,2}$, Qiang Liu$^{1,2}$ and Xing Fu$^{1,2}$}
\affiliation{
	{$^{1}$Key Laboratory of Photonic Control Technology (Tsinghua University), Ministry of Education, Beijing 100084, China}\\{$^{2}$State Key Laboratory of Precision Measurement Technology and Instruments, Department of Precision Instrument, Tsinghua University, Beijing 100084, China}\\{$^{3}$Optoelectronics Research center, University of Southampton, Southampton SO17 1BJ, UK}
}
	
\date{\today}
	
\begin{abstract}
\noindent  \YS{Spatial mode (de)multiplexing of orbital angular momentum (OAM) beams is a promising solution to address future bandwidth issues, but the rapidly increasing divergence with the mode order severely limits the practically addressable number of OAM modes. Here we present a set of multi-vortex geometric beams (MVGBs) as high-dimensional information carriers, by virtue of three independent degrees of freedom (DoFs) including central OAM, sub-beam OAM, and coherent-state phase. The novel modal basis set has high divergence degeneracy, and highly consistent propagation behaviors among all spatial modes, capable of increasing the addressable spatial channels by two orders of magnitude than OAM basis as predicted. We experimentally realize the tri-DoF MVGB mode (de)multiplexing and shift keying encoding/decoding by the conjugated modulation method, demonstrating ultra-low bit error rates (BERs) caused by center offset and coherent background noise. Our work provides a useful basis for next generation of large-scale dense data communication.}
 
		
\end{abstract}

\maketitle

\noindent Multiplexing of independent optical degrees of freedom (DoFs) such as polarization and wavelength have long been implemented to increase the capacity of optical communication systems~\cite{kaminow1996wideband, hung2003optical,chow2004optical, mukherjee2006optical}. The exploration of spatial DoFs of optical fields has offered new possibilities that mode-division-multiplexing (MDM) scales the capacity by a factor equal to the number of spatial modes acting as independent information channel carriers~\cite{gibson2004free, trichili2019communicating, bozinovic2013terabit, zhang2016mode, wang2016advances}. Among all spatial modes, the use of orbital angular momentum (OAM) beams, which can accommodate theoretically infinite orthogonal modes, has generated widespread and significant interest in the last decade~\cite{allen1992orbital,2012Quantum, 2019Experimental,otte2020high,fang2020orbital,qiao2020multi, 2001entanglement, erhard2018twisted, erhard2020advances}.  However, in practice, OAM modal basis set alone cannot reach the capacity limit of a communication channel~\cite{zhao2015capacity}, since the beam diverges rapidly as the OAM order enlarges, which gives rise to increased power loss for a limited-size receiver aperture. To guarantee sufficient received optical power for data recovery, the number of OAM modes that can be practically supported is severely limited under 60~\cite{fu2019demonstration,wang2014n,wang2015ultra}, mostly under 20~\cite{Lei2015Massive,wang2012terabit,krenn2014communication,wang2011high,krenn2016twisted,ren2016experimental,huang2014100}. One can relax the limit and increase the maximum number of addressable spatial channels, by enhancing the divergence degeneracy, i.e. having more orthogonal spatial modes propagating in identical manner. To this end, incorporating both the radial and azimuthal components of Laguerre–Gaussian (LG) beams~\cite{li2017power, guo2018orbital,pang2018400,xie2016experimental} makes one constructive step, but the improvement is far from satisfactory. To meet the growing demand for data capacity, it is highly desirable to use a large set of spatial modes with the variation in beam quality among all modes as small as possible. 

In recent years, a class of exotic structured optical fields termed ray-wave geometric beams has attracted much attention, whereby crafted spatial modes appear to be both wave-like and ray-like ~\cite{buvzek1989generalized,wodkiewicz1985coherent,shen2018periodic, shen2018polygonal, shen2018truncated, lu2011generation, tuan2018characterization, chen2006devil, lu2008three, chen2010spatial, Wan:20}. In the wave picture, the beam is a coherent laser mode, imbued with typical OAM feature. In the ray picture, the mode is coupled with a cluster of geometric rays, unveiling new controllable DoFs that notably increase the divergence degeneracy, such as sub-OAM (partial vortex along each ray) and coherent-state phase (the phase to tune the ray sequence). 

In this work, we demonstrate that the modal basis of ray-wave geometric beams outperforms the OAM and LG modal basis, in terms of approaching the capacity limit of a communication channel. Specifically, we create a three-dimensional set of orthogonal data-carrying beams, by employing three independent intrinsic DoFs of the multi-vortex geometric beam (MVGB), one type of ray-wave geometric beams, including the central OAM, sub-beam OAM and coherent-state phase. 
We show the MVGB set is extremely densely packed in beam quality space, that it can possess a divergence degeneracy as high as 20, a 20X increase over OAM modal basis, and a divergence variation by merely 18$\%$ among 100 independent lowest order spatially multiplexed modes, in contrast to 900$\%$ for OAM counterpart and 429$\%$ for LG counterpart. As a result, thousands of independently spatial channels in MVGB basis can be supported in a free space optical communication system, two orders of magnitude larger than that in OAM basis. To validate the performance of the high dimensional information carriers, we analyze in detail the orthogonality of MVGB mode space on different spatial indices, based on which we experimentally realize the tri-DoF mode (de)multiplexing and shift keying encoding/decoding by the conjugated modulation using digital micro-mirror device (DMD). The results indicate another distinct advantage of MVGB basis in demultiplexing with much lower bit error rate (BER) caused by the center offset and the coherent background noise, comparing with OAM basis, in the conjugated demultiplexing process. We believe the divergence-degenerated MVGB modal set provide a useful basis for boosting the capacity of future optical communication systems.

\begin{figure*}
\centering
\includegraphics[width=\linewidth]{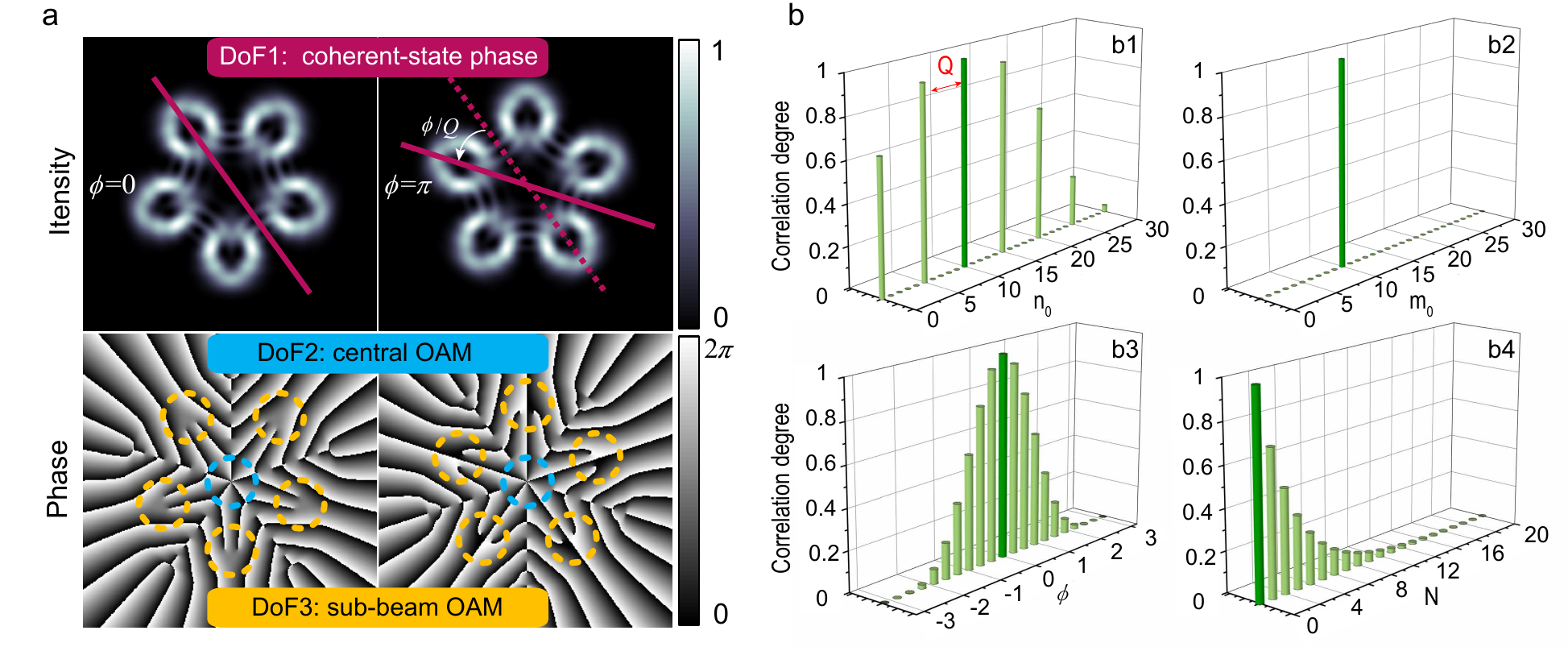}
\caption{(a) Diagram of three DoFs of MVGB ${|\Psi_{7,2}^{(\pi/2,\pi/2,\phi)}\rangle}_{5,0}^{5}$.
(b) correlation degree analysis of MVGBs.
(b1) correlation degree between ${|\Psi_{10,0}^{0}\rangle}^{5}$ and ${|\Psi_{n_{0},0}^{0}\rangle}^{5}$ where $n_{0}$ is changing from 0 to 30,
(b2) correlation degree between ${|\Psi_{5,10}^{0}\rangle}^{5}$ and ${|\Psi_{5,m_0}^{0}\rangle}^{5}$ where $m_{0}$ is changing from 0 to 25,
(b3) correlation degree between ${|\Psi_{5,10}^{0}\rangle}^{5}$ and${|\Psi_{5,10}^{\phi}\rangle}^{5}$ where $\phi$ is changing from $-\pi$ to $\pi$,
(b4) correlation degree between ${|\Psi_{5,10}^{0}\rangle}^{5}$ and ${|\Psi_{5,10}^{\pi/2}\rangle}^{N}$ where $N$ is changing from 0 to 20. The dark bar in each subfigure indicates the reference index value for the correlation degree analysis.}
\label{f.Tri_Dof}
\end{figure*}

\vspace{0.5cm}
\noindent {\textbf{\large{Results}}} \par
\noindent \textbf{\YS{Tri-DoF MVGBs.}} 
The ray-wave geometric beam can be represented as the superposition of a family of eigenstates (Hermite-Laguerre-Gaussian (HLG) modes) with sub-Poissionian distribution:
\begin{equation}
{|\Psi_{n_{0},m_{0}}^{(\alpha,\beta,\phi)}\rangle}_{p,q}^{N} = \frac{1}{2^{N/2}} \sum_{K=0}^N{\left( \begin{matrix}
	N  \\
	K  \\
	\end{matrix} \right)}^{{1}/{2}}{\text{e}}^{\text{i}K{\phi}}\text{HLG}_{n_{0}+pK,m_{0}+qK}^{\left( {\alpha ,\beta } \right)}
\label{su2.wave}
\end{equation}
where $N+1$ is the number of eigenmodes in the frequency-degenerate family of $\text{HLG}_{n_{0}+pK,m_{0}+qK}^{\left( {\alpha ,\beta } \right)}$, $p$ and $q$ are ratios of transverse frequency spaces in $x$- and $y$-axis respectively, $n_0$ and $m_0$ are the initial orders of transverse mode in $x$- and $y$-axis respectively, $l_{0}$ is the initial order of longitudinal mode in $z$-axis, and $\phi$ is the coherent-state phase. In particular, when $\alpha = \beta =\pm\pi/2$, the HLG eigenmode degenerates to LG mode~\cite{Wan:20}. 

For the case of $p=Q$ and $q=0$, the ray-wave geometric beam is referred to as MVGB, having $Q$ vortex sub-beams. In a MVGB, the coherent-state phase $\phi$ acting as one DoF is manifested in the orientation of petal-like intensity pattern, as shown in Fig.~\ref{f.Tri_Dof}(a) that the rotation of orientation angle relative to the case of $\phi=0$ is $\phi/Q$. The other two DoFs to be exploited are $n_0$ and $m_0$, the values of central OAM and sub-beam OAM, respectively, as demonstrated in the phase distribution in Fig.~\ref{f.Tri_Dof}(a). Hereinafter, we focus on the MVGB with $Q$ = 5 as an example, the expression of which is thus abbreviated as ${|\Psi_{n_{0},m_{0}}^{\phi}\rangle}^{N}$ for the sake of brevity.

\begin{figure*}
\centering
\includegraphics[width=\linewidth]{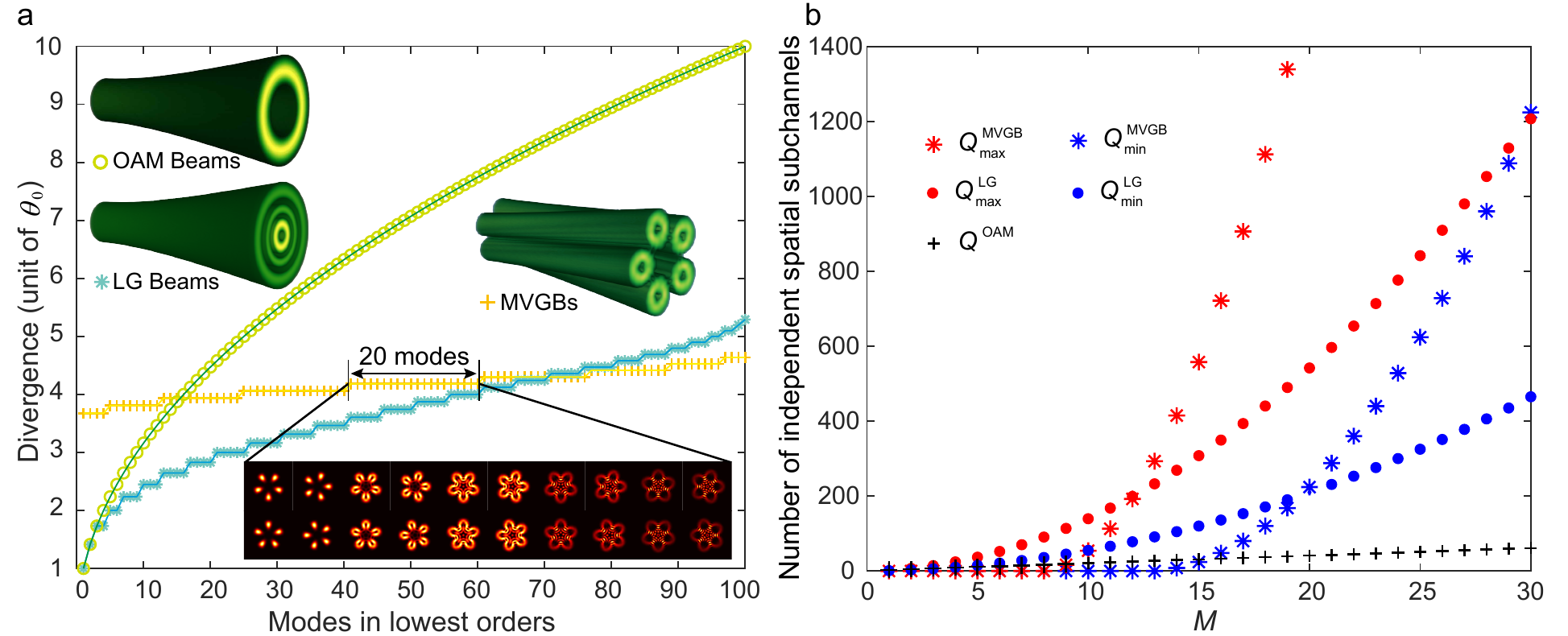}
\caption{(a) Comparison on divergences of MVGBs, OAM modes and LG modes, in respective 100 lowest orders. Insets illustrate three-dimensional diagrams of three type of modes, as well as the transverse intensity profiles of 20 MVGB modes in degenerated divergence; (b) comparison on predicted numbers of independently addressable spatial subchannels in a spatial multiplexing system using MVGBs, OAM modes and LG modes.}
\label{f.Three_Dof&channelnum}
\end{figure*}

Before the demonstration of potential use of MVGBs in MDM communication application, it is crucial to first investigate the orthogonality of MVGBs in terms of three spatial indices, using the correlation degree as the metric. The correlation degree of 0 and 1 represent orthogonal and completely non-orthogonal conditions for two MVGBs, respectively. The correlation degree of two RWGBs with different parameters $n_0$, $m_0$ and $\phi$, as represented by the inner product mathematically, is given by:

\begin{equation}\begin{aligned}
&{}^{{N}}\left \langle \Psi_{n_{0_s},m_{0_s}}^{\phi_s} |\Psi_{n_{0_r},m_{0_r}}^{\phi_r}\right \rangle{}^{{N}}\\
&=\iint\frac{1}{2^{N/2}} \sum_{J=0}^{N}{\left(\begin{matrix}
	N      \\
	J      \\               
	\end{matrix} \right)}^{{1}/{2}}{\text{e}}^{\text{-i}J{\phi_{s}}}\widetilde{\text{HLG}}_{n_{0_s}+pJ,m_{0_s}+qJ}^{\left( {\pi/2,\pi/2 }\right)} \\
&\times {\frac{1}{2^{N/2}}}\sum_{K=0}^{N}{\left( \begin{matrix}
	N      \\
	K      \\
	\end{matrix} \right)}^{{1}/{2}}{\text{e}}^{\text{i}K{\phi_{r}}}\text{HLG}_{n_{0_r}+pK,m_{0_r}+qK}^{\left( {\pi/2,\pi/2} \right)}
	dxdy
    \label{e.orth}
\end{aligned}\end{equation}
where sign '$\sim$' means conjugate, and $\iint{\widetilde{\text{HLG}}_{n_{s},m_{s}}^{\left({\pi/2,\pi/2} \right)}\times\text{HLG}_{n_{r},m_{r}}^{\left( {\pi/2,\pi/2} \right)}dxdy} =\delta_{(n_{s},n_{r}),(m_{s},m_{r})}$ (where $\delta_{(n_{s},n_{r}),(m_{s},m_{r})}\neq0$ only when $n_{s}=n_{r}$ and $m_{s}=m_{r}$). 

According to Eq.~\ref{e.orth}, the theoretical results of the correlation degree analysis of MVGBs are shown in Fig.~\ref{f.Tri_Dof}(b). Basically, three spatial indices ($n_0$, $m_0$ and $\phi$) are uncoupled and independent of each other. The orthogonalities of two MVGBs associated with each of three spatial indices are examined as follows. First, two MVGBs with different central OAM values ($n_0$) are mutually orthogonal to each other when $ \left|n_{{0_s}}-n_{{0_r}}\right|\neq ZQ$ ($Z$ is an integer and $Z \leqslant N$), as shown in Fig.~\ref{f.Tri_Dof}(b1). Second, since $q=0$ for MVGBs, it is natural that MVGBs with different sub-beam OAM values ($m_0$) are orthogonal to each other, same as the case of general OAM beams, as illustrated in Fig.~\ref{f.Tri_Dof}(b2). Third, two MVGBs with the coherent-state phase ($\phi$) being 0 and $\pi$ respectively are orthogonal, which is analogous to the case of left- and right-hand circular polarization states, as depicted in Fig.~\ref{f.Tri_Dof}(b3). Furthermore, note that two MVGBs can be regarded as quasi-orthogonal when $|\phi_s-\phi_r|=\pi/2$ ($N>5$), since the correlation degree is less than 0.1, as described in Fig.~\ref{f.Tri_Dof}(b4). Therefore, $\phi$ can take up to four values $(0,\pi/2,\pi,$ and $3\pi/2)$ when $N>5$, for the realization of efficient mode (de)multiplexing. So far, we have obtained the MVGB modal basis set characterized by three spatial indices of $n_0$, $m_0$ and $\phi$, enabling a combination of $4\times K_{n}\times K_{m}$ readily available spatial modes as information carriers, where $K_{n}$ and $K_{m}$ are the numbers of central OAM and sub-beam OAM states selected from theoretically unbounded states respectively. The detailed orthogonality analysis of general ray-wave geometric beam can be found in Supplementary Note 1.



\vspace{0.5cm}
\noindent \textbf{High divergence degeneracy.}
The beam propagation dynamics in free space is vital for free space optical communication~\cite{forbes2014laser} and governed by the beam quality factor $M^2$ entirely. The dynamic transmission characteristics of the beam include the beam size and divergence angle, where the divergence angle of the beam determines the transverse spatial frequency of the beam. For a $\text{LG}_{pl}$ mode and a MVGB mode as a superposition of multiple higher-order eigenmodes (${\rm{HLG}}_{n,m}^{\left({\alpha,\beta}\right)}$), the beam quality factors are respectively expressed as~\cite{siegman1990new}:

\begin{equation}
 {M_{\rm{LG}}^{2}}=2p+\left |l  \right |+1
\end{equation}

\begin{equation}
 {M_{\rm{MVGB}}^2}{\rm{ = }}\sum\limits_{m = 0}^\infty{\sum\limits_{n = 0}^\infty {\left( {n+ m + 1} \right){{\left| {{c_{nm}}} \right|}^2}} }
 \label{e.M2}
\end{equation}
where $c_{nm}=\frac{1}{2^{N/2}} {\left( \begin{matrix}
	N  \\
	K  \\
	\end{matrix} \right)}^{{1}/{2}}$ are normalized amplitudes for the eigenmodes $\text{HLG}_{n_{0}+QK,m_{0}}^{\left( {\alpha ,\beta } \right)}$ with $n=n_{0}+QK$ and $m=m_{0}$ in Eq.~\ref{su2.wave}, and the total power $\sum\limits_{m = 0}^\infty  {\sum\limits_{n = 0}^\infty  {{{\left| {{c_{nm}}} \right|}^2}} }= \frac{1}{2^{N/2}} \sum_{K=0}^N{\left( \begin{matrix}
	N  \\
	K  \\
	\end{matrix} \right)}^{{1}/{2}}= 1$.

It can be seen from Eq.~\ref{e.M2} that the beam quality factor of the MVGB depends entirely on the family of eigenmodes it contains and the corresponding normalized weighting factor. Notably, the coherent-state phase parameter does not affect the superposition components and weighting factors, thus the additional DoF of $\phi$ can scale the beam quality degeneracy by a factor of 4 (equal to the number of employed values of $\phi$), compared with two-dimensional LG modal basis. For instance, as Supplementary Fig. 3 shows, among the 100 lowest orders of MVGB modes, by combinations of $m_0$=$\{0, 1, 2, 3, 4\}$, $n_0$=$\{0, 1, 2, 3, 4\}$, and $\phi$=$\{0,\pi/2,\pi,3\pi/2\}$, the maximum beam quality degeneracy reaches as high as 20, that up to 20 modes share the same beam quality factor of $M^2$=17.5. This leads to only a total of 9 beam quality factors from all the 100 modes: $M^2$=$\{13.5, 14.5, 15.5, 16.5, 17.5, 18.5, 19.5, 20.5, 21.5\}$. In contrast, the 100 lowest orders of $\text{LG}_{pl}$ modes, by combinations of $p$ and $l$ both taking 10 integer values from 0 to 9, have 28 integer values of $M^2$ from 1 to 28. 

Similarly, the divergence and beam waist diameter of MVGBs have high degeneracies. The beam size can be calculated by the second moment of intensity~\cite{willner2016design}, thus we have the beam waist diameter $D_{0,m}$ and divergence $\theta_m$ of m-order LG modes and MVGBs expressed as:
 
\begin{equation}\begin{aligned}
 D_{0,m}&=2\sqrt \frac{{2\int_0^{2\pi}{\int_0^\infty{{r^2}I_0(r,\phi)rdrd\phi}}}}{{\int_0^{2\pi}{\int_0^\infty{I_0(r,\phi)rdrd\phi}}}} \\
\theta_m&=\frac{M^2\lambda}{\pi D_{0,m}}
\end{aligned}\end{equation} 
where $I_0(r,\phi)$ is the intensity distribution of beam waist cross-section. 

Figure~\ref{f.Three_Dof&channelnum}(a) compares the divergences of OAM modes, LG modes and MVGBs in respective 100 lowest orders, all normalized to the divergence angle of fundamental Gaussian beam ($\theta_0$). Note that the maximum divergence degeneracy is 20, that 20 modes share the same divergence of $\theta_m$=4.18 $\theta_0$, and the divergence varies by merely 18$\%$ among 100 independent lowest order spatially multiplexed modes of MVGBs, in contrast to 900$\%$ for OAM counterpart and 429$\%$ for LG counterpart. This brilliant feature of MVGB basis results in a highly consistent propagation behavior of data channels, which is beneficial for the beam tracking, and alignment control of receiver optics and adaptive optics~\cite{trichili2019communicating}. The corresponding variations in beam waist diameter for LG modes and MVGBs are compared in Supplementary Figure 4.

By the virtue of high divergence degeneracy, we claim that the MVGB basis can achieve capacity beyond OAM and LG counterparts. To confirm this, we count the number of MVGB modes that fit into a line-of-sight free space communication system with a space–bandwidth product (SBP) of $2{R_0} \times 2{\rm{NA}/\lambda}$, where $R_0$ and $\rm{NA}$ are the aperture radius and numerical aperture of both circular apertures of transmitter and receiver, and $\lambda$ is the wavelength. Following the procedure of ~\cite{zhao2015capacity}, we define a dimensionless parameter $M$, which is $\pi/4$ times the $\rm{SBP}$, and then estimate the lower and upper bound on the number of independently addressable spatial subchannels $Q$, counting all MVGB modes that satisfy $M_{\rm{MVGB}}^2\leqslant$M, as given by

\begin{equation}\begin{aligned}
Q_{\min }^{\text{MVGB}}&\text{=}8\times \frac{\lfloor M-12.5\rfloor \times \left(\lfloor M-12.5\rfloor+1 \right)}{2}\\
&\ge 4\left( M-13 \right)\left( M-12 \right)
\label{de.lower}
\end{aligned}\end{equation}
where $\lfloor \rfloor$ represents the floor integration.

\begin{equation}\begin{aligned}
Q_{\max }^{\text{MVGB}}\ge 4\left( \frac{16M}{{{\pi }^{2}}}-13 \right)\left( \frac{16M}{{{\pi }^{2}}}-12 \right)
\label{de.upper}
\end{aligned}\end{equation}

The estimated subchannel numbers of MVGB multiplexing based on Eq.~\ref{de.lower} and Eq.~\ref{de.upper} are compared with those of OAM multiplexing and LG mode multiplexing in Fig.~\ref{f.Three_Dof&channelnum}(b). It is noteworthy that although the lowest beam quality of MVGB is higher than OAM beams and LG modes, corresponding to a larger intercept in the x-axis, the lower and upper bound curves of MVGB have far steeper slopes versus $M$ (equivalently the SBP) than OAM and LG modal basis, thereby accomodating far more data channels. For example, at $M$=30 (SBP=38), $Q^{\rm{OAM}}$ is about 60, $Q^{\rm{LG}}$ lies between 460 and 1200, and $Q^{\rm{MVGB}}$ lies between 1200 and 5000, which is about two orders of magnitude larger than OAM basis. The superiority of MVGB modal basis would become even more distinct, in terms of addressable independent spatial subchannels, when the communication system has higher SBP and larger scales, as indicated by the trend in Fig.~\ref{f.Three_Dof&channelnum}(b). 

\begin{figure*}
\centering
\includegraphics[width=0.78\linewidth]{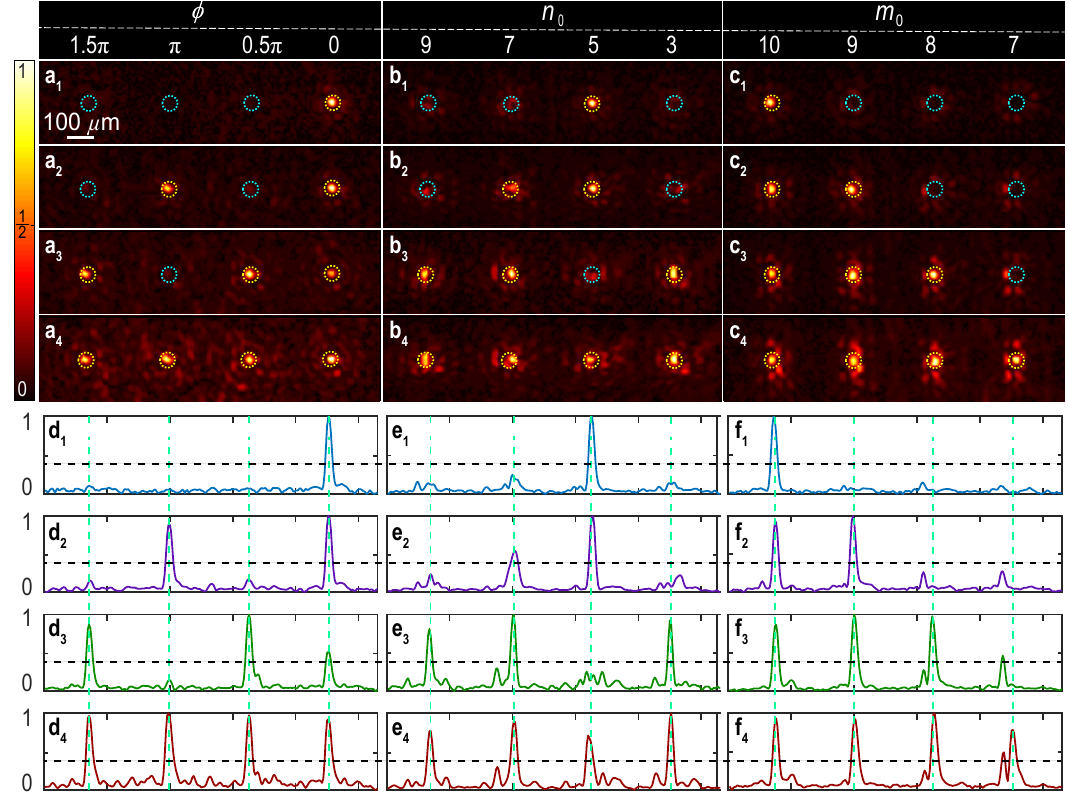}
\caption{Tri-DoF demultiplexing results of MVGBs. Three demultiplexed conjugated holographic masks for DoFs $\phi,n_0,m_0$ are obtained from non-collinear superposed conjugated optical field of MVGBs: $\sum_{\phi=0,\pi/2,\pi,3\pi/2}{|\widetilde{\Psi}_{5,10}^{\phi}\rangle^{5}}\left(\text{i}2\pi {{u}_{\phi}}x \right)$, $\sum_{n_{0}=3,5,7,9}{{|\widetilde\Psi}_{n_{0},10}^{0}\rangle^{5}} \left(\text{i}2\pi {{u}_{n_0}}x \right)$, and $\sum_{m_{0}=7}^{10}{|\widetilde{\Psi}_{5,m_0}^{0}\rangle^{5}} \exp\left(\text{i}2\pi {{u}_{m_0}}x \right)$. $(\text{a}_{1 \text{–} 4}) \text{ to } (\text{c}_{1 \text{–} 4})$ are experimental intensity profiles of demultiplexed beam components for incident light with one to four collinear superposed MVGB, respectively.  $(\text{d}_{1 \text{–} 4})\text{ to }(\text{f}_{1 \text{–} 4})$ are corresponding one dimensional cross-section views of $(\text{a}_{1 \text{–} 4}) \text{ to } (\text{c}_{1 \text{–} 4})$. Black dash line: discrimination threshold.}
\label{f.RWD Multiplexing}
\end{figure*}

\begin{figure*}[t!]
\centering
\includegraphics[width=0.78\linewidth]{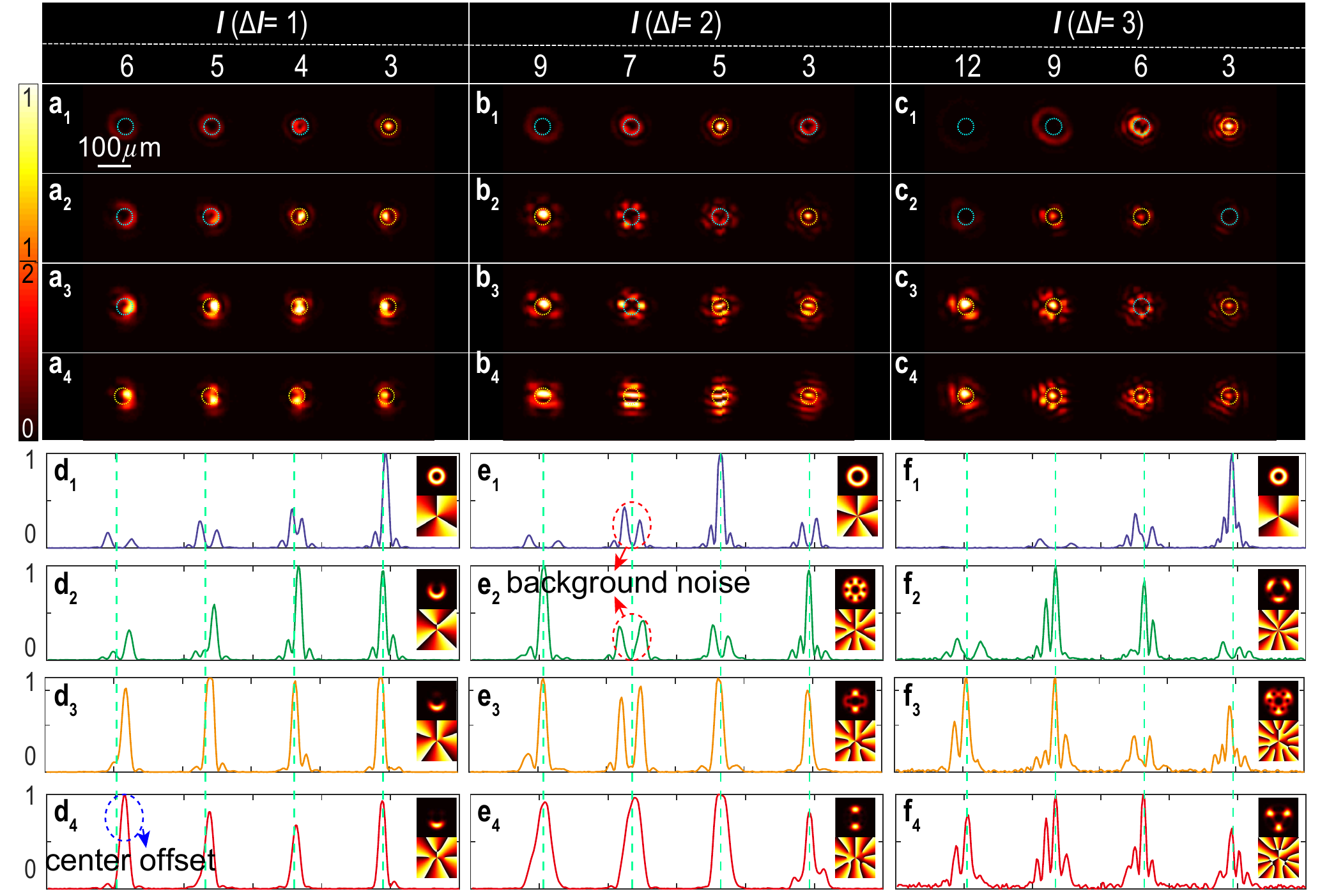}
\caption{Demultiplexing results of OAM beams. Three demultiplexed conjugated holographic masks for OAM mode spacing of 1, 2 and 3 are obtained from non-collinear superposed conjugated optical field of MVGBs: $\sum_{l=3}^6\widetilde{\text{LG}}_{p=0}^{l}\left(\text{i}2\pi {{u}_{l}}x \right)$, $\sum_{l=3,5,7,9}\widetilde{\text{LG}}_{p=0}^{l}\left(\text{i}2\pi {{u}_{l}}x \right)$, and $\sum_{l=3,6,9,12}\widetilde{\text{LG}}_{p=0}^{l}\left(\text{i}2\pi {{u}_{l}}x \right)$. $(\text{a}_{1 \text{–} 4}) \text{ to } (\text{c}_{1 \text{–} 4})$ are experimental intensity profiles of demultiplexed beam components for incident light containing one to four collinear superposed OAM beams, respectively.  $(\text{d}_{1  \text{–}  4})\text{ to }(\text{f}_{1 \text{–} 4})$ are corresponding one dimensional cross-section view of $(\text{a}_{1 \text{–} 4}) \text{ to } (\text{c}_{1 \text{–} 4})$, where the insets are the intensity and phase distribution of the corresponding input field.}
\label{f.OAM Multiplexing}
\end{figure*}

\vspace{0.5cm}
\noindent \textbf{\YS{Low BER in MVGB demultiplexing.}}
Despite of the widely applied sorting approaches developed for OAM beams and LG beams, such as Dammann vortex grating~\cite{Lei2015Massive,Zhang2010Analysis,PhysRevApplied.12.064062}, Gouy phase radial mode sorter~\cite{gu2018gouy}, log-polar based azimuthal mode sorters~\cite{PhysRevLett.105.153601, Mirhosseini2013Efficient}, interference and diffraction method~\cite{PhysRevLett.88.257901,leach2004interferometric,hickmann2010unveiling}, and deep learning\cite{PhysRevLett.123.183902}, identification and sorting of ray-wave geometric beams such as MVGBs is still at its infancy, due to the intrinsic complex structure and rich controlling parameters. Inspired by the recent work of digital cavity-free tailoring~\cite{Wan:20}, here we demonstrate the sorting and demultiplexing of superposed MVGBs in the following experiments, using the demultiplexed conjugated holographic masks that are designed by diffracting each beam component into different location~\cite{2019Structured,2019A}, as detailed in the Methods section.

Figure~\ref{f.RWD Multiplexing} shows the experimental results of MVGB demultiplexing associated with each of three spatial indices. For the DoF of $\phi$, subfigures $\text{a}_{1 \text{–} 4}$ demonstrate the intensity profiles of demultiplexed beam components separated along the $x$ direction. The corresponding input collinear superposed MVGBs containing one to four beam components are $\sum_{\phi=0}{|\Psi_{5,10}^{\phi}\rangle}^{5}$, $\sum_{\phi=0,\pi}{|\Psi_{5,10}^{\phi}\rangle}^{5}$, $\sum_{\phi=0,\pi/2,3\pi/2}{|\Psi_{5,10}^{\phi}\rangle}^{5}$, and $\sum_{\phi=0,\pi/2,\pi,3\pi/2}{|\Psi_{5,10}^{\phi}\rangle}^{5}$, respectively. Corresponding demultiplexed conjugated holographic masks are designed as $T(x,y)=\frac{1}{2}+\frac{1}{2}\text{sign}\left[ \cos \left(\Phi\right )+\cos\left(\text{arcsin}A \right)\right]$, where $A$ and $\Phi$ are respectively the amplitude and phase of non-collinear superposed conjugated optical field of MVGBs $\sum_{\phi=0,\pi/2,\pi,3\pi/2}{|\widetilde{\Psi}_{5,10}^{\phi}\rangle}^{5} \exp\left(\text{i}2\pi {{u}_{\phi}}x \right)$ (see details in the Methods section). The four dotted circles in each subfigure of $\text{a}_{1 \text{–} 4}$ indicate the target diffracting locations of all four beam components by the holographic mask design, among which the yellow ones are signal locations, corresponding to those beam components that are practically multiplexed in the input beams, and the blue ones are non-signal locations. 
The same method is applied to OAM demultiplexing for the comparison in the next subsection, as shown in Fig.~\ref{f.OAM Multiplexing}. Moreover, an 8-bit and 16-bit hybrid shift keying encoding/decoding scheme with tri-DoF MVGBs are demonstrated with zero BER, as detailed in Supplementary Note 3.

Another advantage of MVGB multiplexing is manifested in the low BER in the demultiplexing process of conjugated modulation. Coherent light sources are widely exploited in MDM communication, since the mode coding can be done by a modulation device and does not require the spatial coupling. However, coherence may bring difficulty to signal decoding and thus increase the BER, which is defined as the proportion of bit values that are incorrectly identified according to the demultiplexing results. On one hand, the intensity peak of a coherently superposed beam may deviate from the copropagating optical axis, leading to the center offset, lateral displacement of intensity peak relative to the target location, of decoding spot and thus possibly introducing a bit error that the bit value of 1 is incorrectly identified as 0, e.g. the marked peak in Fig.~\ref{f.OAM Multiplexing} ($\text{d}_4$) for the case of OAM beams. The BER caused by center offset depends on the radius of discrimination region (RD) that the bit error induced by center offset is valid only when the offset is larger than the RD. On the other hand, the background noise embedded in different beam components may be coherently superposed near the target diffracting position, resulting in a noise peak beyond the DT that may introduce a bit error that bit value of 0 is incorrectly identified as 1, e.g. the marked peaks in subfigures $\text{e}_{1}$ and $\text{e}_{2}$ of Fig.~\ref{f.OAM Multiplexing}. The BER caused by background noise depends on DT that a bit error is valid when the intensity peak of background noise is higher than DT in the discrimination region. 

\begin{figure*}[t!]
\centering
\includegraphics[width=\linewidth]{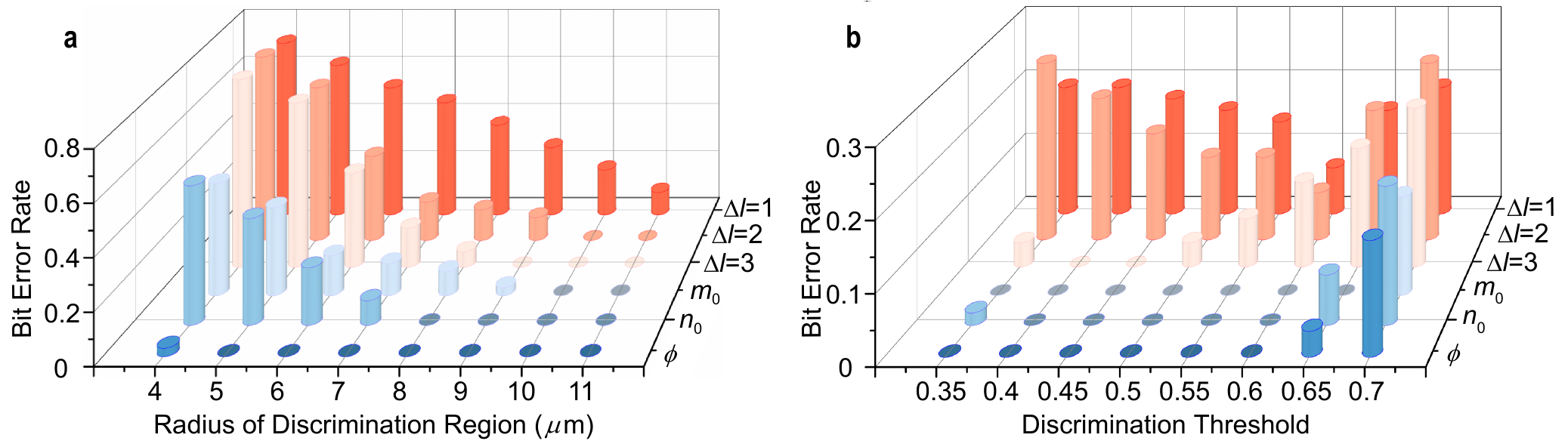}
\caption{The BER caused by center offset (a) and background noise (b) in the demultiplexing: comparison among the OAM beams with different mode spacings and tri-DoF MVGBs.}
\label{f.Error}
\end{figure*}

We emphasize that MVGBs have remarkably stronger capability, in contrast to OAM beams, in suppressing the BER caused by center offset and non-negligible background noise in the demultiplexing process of coherent light, thereby efficiently improving the signal-to-noise ratio and reducing the inter-modal crosstalk. First, due to the inherently more complicated intensity and phase distribution of MVGB than OAM beam, the intensity profile of superposed MVGBs can well maintain the centrosymmetric feature, which is vital in mitigating the center offset. Second, for the input beam components that do not match with the demultiplexed mask, the complex phase structure of MVGB enables the responses at the non-signal locations as weak dispersing speckles (see Fig.~\ref{f.RWD Multiplexing}), rather than concentrated spot patterns which are common in the case of OAM beams (see Fig.~\ref{f.OAM Multiplexing}). 

To further verify the superiority, we compare the demultiplexing performance of MVGBs and OAM beams in terms of the measured BERs caused by center offset and background noise, as shown in Fig.~\ref{f.Error} (a) and (b), respectively. The average offsets of the MVGBs on three DoFs ($\phi, n_0, m_0$) are 2.69 $\mu$m, 4.47 $\mu$m and 3.95 $\mu$m, respectively, while those for OAM beams (mode spacing of $\Delta l$=1, 2, and 3) are 6.62 $\mu$m, 5.31 $\mu$m, and 5.25 $\mu$m, respectively. As a result, the BER averaging among all values of RD (from 4 $\mu$m to 11 $\mu$m) are 0.16, 0.37, and 0.31 for MVGBs ($\phi, n_0, m_0$), and are 	0.59,	0.48, and 0.46 for OAM beams ($\Delta l$=1, 2, and 3), respectively, as shown in Fig.~\ref{f.Error} (a). It is of particular interest that MVGBs on the DoFs of $\phi$ and $n_0$ have zero BER with RD above 5 $\mu$m and 8 $\mu$m, respectively. Figure~\ref{f.Error} (b) shows the BER results merely induced by the background noise, in which a large RD as 70 $\mu$m is used to prevent the influence of center offset on the BER. It is impressive that MVGB has zero BER for demultiplexing on all three DoFs at 0.4 $\leqslant$ DT $\leqslant$ 0.6, in which the lower bound of zero BER for the case of $\phi$ can reach as low as 0.2 (not shown in the plot). In contrast, demultiplexing of OAM beams yields a high BER by background noise, which increase from 0.071 with $DT=0.6$ to 0.196 with $DT=0.4$, for the cases of OAM mode spacing of 1. When the OAM mode spacing increases to 3, the BER reduces to zero but at a narrow DT range (0.4 $\leqslant$ ST $\leqslant$ 0.45). Note that the BER increases rapidly for all the cases with DT higher than 0.6, which is not caused by an extremely high noise level beyond DT, but by the uneven intensity responses among different beam components of signal causing that certain beam component has a smaller signal intensity than DT and the bit value at corresponding channel is incorrectly identified as 0. All these experimental results show that the tri-DoF MVGBs outperform the general OAM beams in terms of low BER, indicating that tri-DoF MVGBs are advantageous as potential high-dimensional information carriers.


\vspace{0.5cm}
\noindent{\bf{\large{Discussion}}} \par
The emergence of structured light offers a possible solution to meet future demands for communication capacity, by utilizing its spatial DoFs from high-dimensional orthogonality. In this work, we introduce a novel modal basis of MVGBs with three DoFs including the central OAM, sub-beam OAM and coherent-state phase. At the heart of our work is the exploitation of modes in ray-wave duality state, which allows us access to higher divergence degeneracy and more consistent propagation behavior among all modes, dramatically increasing the addressable number of independent spatial channels. We validate the potentials of spatially multiplexed MVGB as high dimensional information carriers, by proof-of-concept experiments of the tri-DoF mode (de)multiplexing and shift keying encoding/decoding. Notably, we make the challenging demultiplexing task of tri-DoF modes possible, by proposing a novel approach based on conjugated modulation that can fully resolve the ray-wave duality state, removing the long-standing obstacle in identification and sorting of ray-wave geometric beam that has prohibited its progress. The decoding results show that MVGB modal basis has significant strengths in suppressing BERs induced by center offset and coherent background noise. 

The MVGB multiplexing is compatible with and can combine with other techniques, such as wavelength and polarization division multiplexing, to further increase capacity and facilitate the implementation of next generation high-capacity communication network. Our technique could be extended to other types of ray-wave geometric beams, to explore even more spatial DoFs and higher divergence degeneracy. The concept of tri-DoF modal basis can also be applied to encoding and decoding in the quantum data channels. In the future, we will exploit the indivisibility among multiple DoFs to further explore its value in realizing high-dimensional multi-partite entanglement.

\vspace{0.5cm}
\noindent{\bf{\large{Methods}}} \par
\begin{figure}[h!]
\centering
\includegraphics[width=\linewidth]{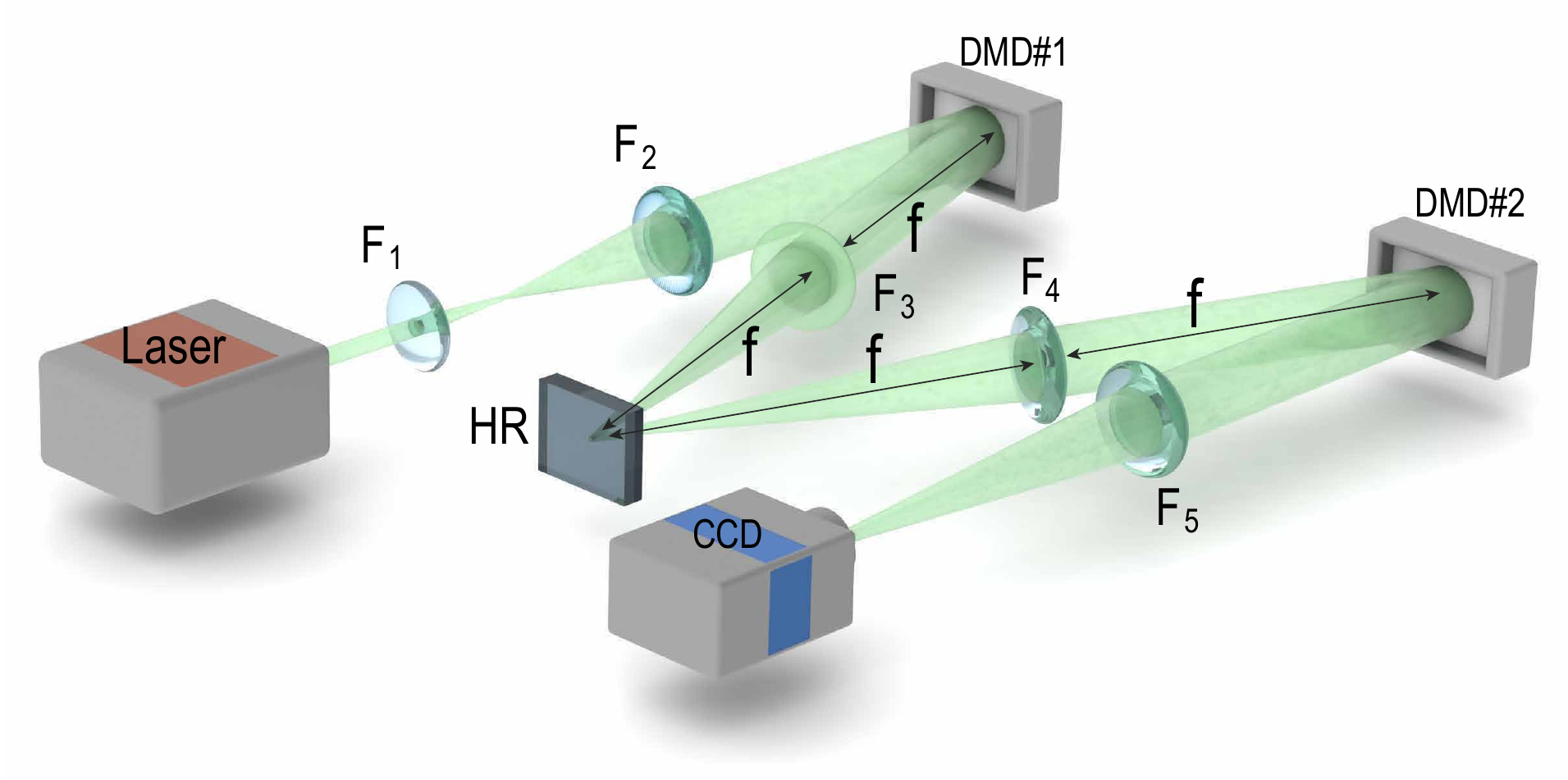}
\caption{Schematic of experimental setup, $\text{F}_1-\text{F}_5$: lens, HR: High reflective mirror; CCD: charge coupled device (Microview RS-A1500-GM60 with a resolution of $1280 \times 1024$ pixels and a pixel size of 5.3 $\mu$m), DMD: digital micro-mirror device (F6500 Type A 1080P VIS KIT, resolution: 1920×1080 pixels).}
\label{f.experiment setup}
\end{figure}

\noindent \textbf{Experimental setup.}
\label{Experiment}
The experimental setup is shown in Fig.~\ref{f.experiment setup}. The beam from a solid laser source (CNI laser, MGL-III-532nm) is expanded to a near-plane wave by passing through a telescope ($\text{F}_1$, focal length of 25 mm; $\text{F}_2$, focal length of 300 mm) with the magnification of 1:12, and illuminates DMD $\#$1 loaded with a hologram of the target light field. Then the first order of the beam is selected and reflected by the HR mirror, image relayed by a $4f$ system with $\text{F}_3$ and $\text{F}_4$, both with the focal length of 150 mm, and transmitted to DMD$\#$2 loaded with holograms of corresponding conjugated optical field. The modulated beam is focused into a spot by a convex lens ($\text{F}_5$, focal length of 150 mm), the intensity profile of which is captured by a CCD camera located at the focal plane of $\text{F}_5$.

\begin{figure}[h!]
\centering
\includegraphics[width=\linewidth]{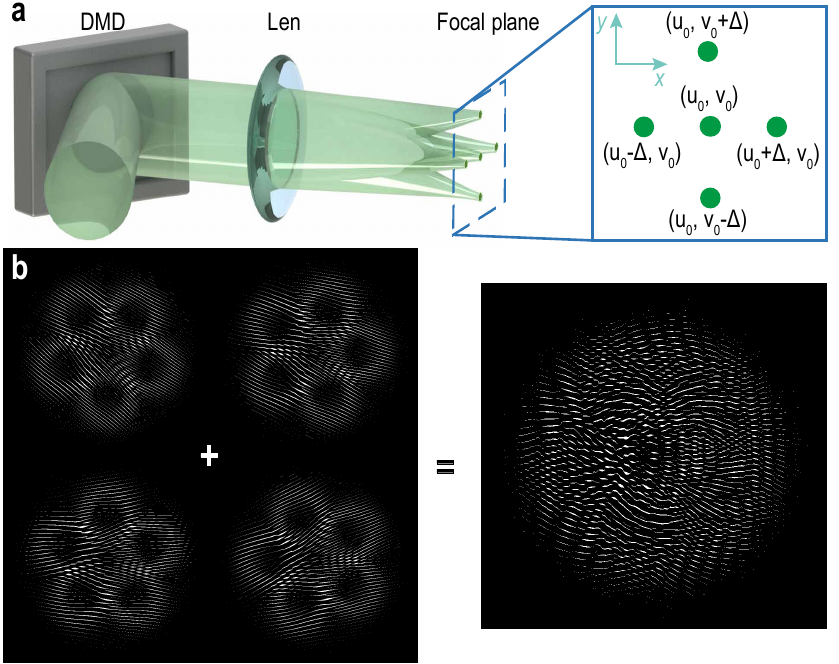}
\caption{(a) Diagram of beam spatial demultiplexing. The identified sub-beam can be spatially separated by adding a different linear grating to the corresponding conjugate optical field. (b) conjugated holographic masks of four identified beam components with different linear grating and demultiplexed conjugated holographic mask.}
\label{f.de.mask}
\end{figure}

The discrimination and measurement of correlation degree of ray-wave geometric beams can be realized by mode projective measurement. The hologram for producing the input beams is loaded into DMD $\#$1, then we load a series of corresponding conjugated holographic mask into DMD $\#$2 sequentially and capture the corresponding focal spots using CCD camera, which performs a modal decomposition on every DoF. The measured results of correlation degree are shown in Supplementary Figure 2. Meanwhile, in the demultiplexing experiments for MVGBs and OAM beams, DMD $\#$1 is loaded with the hologram of the collinearly superposed light field for generating the multiplexed beam, and DMD $\#$2 is loaded with the demultiplexed hologram for separating and identifying beam components in the multiplexed beam.

\vspace{0.5cm}
\noindent \textbf{Demultiplexing design of MVGBs.} 
The DMD transmission function of the hologram of conjugated optical field modulation is given as:
\begin{equation}
{{T}_{s}}\left(x,y\right)=M[A,-\Phi+2\pi \left({{u}_{0}}x+{{v}_{0}}y \right)]
\label{de.mask}
\end{equation}
where $M(\alpha,\beta)=\frac{1}{2}+\frac{1}{2}\text{sign}\left[ \cos \left(\beta\right )+\cos\left(\text{arcsin}\alpha \right)\right]$ (see details in Supplementary Note 2).
The target diffracting position of a beam component is determined by the linear grating period $(u_{0}, v_{0})$. The demultiplexed conjugated holographic mask is calculated by a non-collinear superposed conjugated optical field with different periods of linear grating, which means the diffraction direction of each beam component is separated, as shown in Fig.~\ref{f.de.mask}(a). The non-collinear superposed conjugated optical field of MVGBs is:
\begin{equation}
CSU=\sum_{n}\widetilde{SU}_{s}^{n}=\sum_{n}\widetilde{SU}^{n}\exp\left [\text{i}2\pi\left( {{u}_{n}}x+{{v}_{n}}y \right) \right].
\end{equation}
where $SU^n$ are a set of orthogonal MVGB components. According to Eq.~\ref{de.mask}, different linear grating is added to each conjugate optical field, as described in Fig.~\ref{f.de.mask}(b). The demultiplexed conjugated holographic mask can be obtained as:
\begin{equation}\begin{aligned}
 {{T}_{s}^{D}}\left(x,y\right)=M\left(A^{D}, \Phi^D\right)
\end{aligned}\end{equation}
where $A^{D}$ and $\Phi^D$ are normalized amplitude and phase of $CSU$, respectively. The $u_{n}$ and $v_{n}$ are the reciprocal of the period of linear grating in $x$ and $y$ direction of the $n$-th multiplexed mode, respectively.


\vspace{0.5cm}
\noindent{\bf{\large{Acknowledgements}}} \par
This work was funded by National Natural Science Foundation of China (61975087).
	
\vspace{0.5cm}
\noindent{\bf{\large{Competing financial interests}}} \par
The authors declare no financial or competing interests.

\vspace{0.5cm}
\noindent{\bf{\large{Materials and correspondence}}} \par
Correspondence and requests for materials should be addressed to X.F. (fuxing@mail.tsinghua.edu.cn) and Q.L. (qiangliu@mail.tsinghua.edu.cn).

\bibliography{main}
\end{document}